\documentclass[prl,twocolumn,amsmath,amssymb,superscriptaddress, bibnotes]{revtex4}

\usepackage{color}
\usepackage{graphicx}
\usepackage{dcolumn}
\usepackage{bm}

\begin{document}
\newcommand{\red}[1]{\textcolor{red}{#1}}

\title{Universal relationship between low-energy antiferromagnetic fluctuations and superconductivity in BaFe$_{2}$(As$_{1-x}$P$_{x}$)$_{2}$}

\author{Shunsaku~Kitagawa}
\email{kitagawa.shunsaku.8u@kyoto-u.ac.jp}
\author{Takeshi~Kawamura}
\author{Kenji~Ishida}
\affiliation{Department of Physics, Kyoto University, Kyoto 606-8502, Japan}

\author{Yuta Mizukami}
\affiliation{Department of Advanced Materials Science, University of Tokyo, Kashiwa, Chiba 277-8561, Japan}

\author{Shigeru Kasahara}
\affiliation{Department of Physics, Kyoto University, Kyoto 606-8502, Japan}

\author{Takasada Shibauchi}
\affiliation{Department of Advanced Materials Science, University of Tokyo, Kashiwa, Chiba 277-8561, Japan}

\author{Takahito Terashima}
\author{Yuji Matsuda}
\affiliation{Department of Physics, Kyoto University, Kyoto 606-8502, Japan}

\date{\today}

\begin{abstract}
To identify the key parameter for optimal superconductivity in iron pnictides, we measured the $^{31}$P-NMR relaxation rate on BaFe$_{2}$(As$_{1-x}$P$_{x}$)$_{2}$ ($x = 0.22$ and 0.28) under pressure and compared the effects of chemical substitution and physical pressure. 
For $x = 0.22$, structural and antiferromagnetic (AFM) transition temperatures both show minimal changes with pressure up to 2.4~GPa, whereas the superconducting transition temperature $T_{\rm c}$ increases to twice its former value.
In contrast, for $x=0.28$ near the AFM quantum critical point (QCP), the structural phase transition is quickly suppressed by pressure and $T_{\rm c}$ reaches a maximum. 
The analysis of the temperature-dependent nuclear relaxation rate indicates that these contrasting behaviors can be quantitatively explained by a single curve of the $T_{\rm c}$ dome as a function of Weiss temperature $\theta$, which measures the distance to the QCP. 
Moreover, the $T_{\rm c}$-$\theta$ curve under pressure precisely coincides with that with chemical substitution, which is indicative of the existence of a universal relationship between low-energy AFM fluctuations and superconductivity on BaFe$_{2}$(As$_{1-x}$P$_{x}$)$_{2}$.
\end{abstract}


\maketitle

Identifying the key parameter that determines the optimal superconducting transition temperature ($T_{\rm c}$) in the superconducting phase diagrams involving other electronic orders is of primary importance to understand the mechanism of superconductivity.
In the Bardeen-Cooper-Schrieffer (BCS) theory, $T_{\rm c}$ at the weak coupling limit is expressed as\cite{J.Bardeen_PR_1957}
\begin{align}
T_{\rm c} = \frac{1.13\hslash \omega_{\rm D}}{k_{\rm B}}\exp\left(-\frac{1}{N(0)V}\right),
\end{align}
where $\omega_{\rm D}$ is the Debye frequency, $k_{\rm B}$ is Boltzmann's constant, $N(0)$ is the density of states at the Fermi energy, and $V$ is the pairing electron-phonon interaction.
Therefore, it is well known that the $T_{\rm c}$ of a BCS superconductor is affected by the isotope's mass and pressure, both of which change $\omega_{D}$ and/or $N(0)$.
On the other hand, in materials that exhibit superconductivity in the vicinity of the antiferromagnetic (AFM) order, such as cuprates, iron pnictides, and heavy-fermion  superconductors, it has been pointed out that $T_{\rm c}$ is roughly proportional to the characteristic energy of spin-fluctuations based on self-consistent renormalization (SCR) theory\cite{T.Moriya_JPSJ_1994,Y.Nakai_PRB_2013,J.L.Sarrao_PhysicaC_2015}, suggesting that these superconductors are mediated by the AFM fluctuations.
However, it is not straightforward to find the most significant parameter for optimizing $T_{\rm c}$ even in these superconductors because pressure and chemical substitutions, which are general methods to tune the N\'eel temperature $T_{\rm N}$, and $T_{\rm c}$, also change several physical quantities in these superconductors.

\begin{figure}[!tb]
\vspace*{10pt}
\begin{center}
\includegraphics[width=8.5cm,clip]{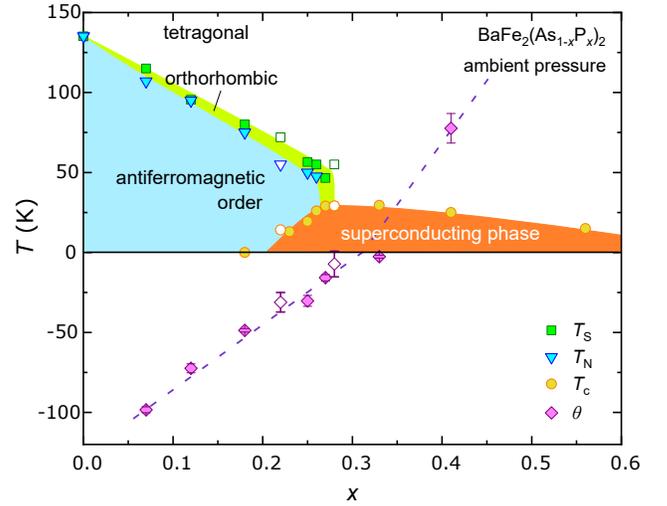}
\end{center}
\caption{(Color online) P concentration $x$ - $T$ phase diagram of BaFe(As$_{1-x}$P$_{x}$)$_{2}$ at ambient pressure\cite{Y.Nakai_PRL_2010}.
Squares, triangles, circles, and diamonds represent the structural phase transition temperature $T_{\rm S}$, N\'eel temperature $T_{\rm N}$, superconducting transition temperature $T_{\rm c}$, and the Weiss temperature $\theta$, respectively.
The open symbols indicate the data from the samples in this study.
The dashed line is intended to guide the eye.
}
\label{Fig.1}
\end{figure}
BaFe$_{2}$(As$_{1-x}$P$_{x}$)$_{2}$, which has a tetragonal ThCr$_{2}$Si$_{2}$-type structure with space group $I4/mmm$ ($D^{17}_{4h}$, No.139), is a member of the iron-based superconductors.
BaFe$_{2}$(As$_{1-x}$P$_{x}$)$_{2}$ is known to be one of the best compounds for investigations among the iron-based superconductors, because superconductivity is induced by the isovalence substitution of P; furthermore, clean single crystals, in which the quantum oscillations are observable, are obtained.    
Figure~\ref{Fig.1} shows the $x$ - $T$ phase diagram of BaFe(As$_{1-x}$P$_{x}$)$_{2}$ as a function of the concentration of P at ambient pressure\cite{Y.Nakai_PRL_2010}.
The resistivity, nuclear magnetic resonance (NMR), and penetration depth measurements indicate that an AFM quantum critical point (QCP) is located at $x \sim 0.3$, and $T_{\rm c}$ reaches a maximum near the QCP\cite{Y.Nakai_PRL_2010,K.Hashimoto_Science_2012}.
According to SCR theory, in the case of a two-dimensional AFM metal, the distance from the QCP can be determined from the Weiss temperature $\theta$, evaluated by fitting the nuclear spin-lattice relaxation rate divided by temperature, $1/T_1T$ to the Curie-Weiss formula\cite{T.Moriya_JPSJ_1990},
\begin{align}
\frac{1}{T_1T} = a + \frac{b}{T+\theta},
\label{eq.2}
\end{align}
where $a$ originates from the intra-band contributions related to the density of states, and $b$ is related to the strength of AFM fluctuations, thus $-\theta$ is regarded as the temperature at which the AFM correlations diverge, i.e., the AFM ordering temperature. 
The sign of $\theta$ is changed by varying the P substitution and $\theta$ becomes zero at $x \sim 0.3$, indicating the existence of an AFM QCP.
A similar relationship between superconductivity and AFM QCP was observed in other ``122'' systems\cite{F.L.Ning_PRL_2010,R.Zhou_NatCommun_2013,G.F.Ji_PRL_2013,M.Miyamoto_PRB_2015}.
In addition, ac susceptibility measurements on BaFe$_{2}$(As$_{1-x}$P$_{x}$)$_{2}$ under pressure revealed that the pressure dependence of $T_{\rm c}$ has a dome shape similar to  the isovalent P substitution phase diagram at ambient pressure\cite{E.Klintberg_JPSJ_2010}.
However, the extent to which AFM fluctuations are changed by pressure and the relationship between AFM fluctuations and superconductivity as a result of the changing pressure has not yet been reported.
In general, an isovalent substitution does not always give the same effect as an applying pressure, e.g., phase diagrams are quite different Fe(Se$_{1-x}$S$_{x}$)\cite{S.Hosoi_PNAS_2016} and pressurized FeSe\cite{J.P.Sun_NatCommun_2016} as well as between Ce(Ir$_{1-x}$Rh$_{x}$)In$_{5}$ and pressurized CeIrIn$_{5}$\cite{S.Kawasaki_PRL_2006}.
In BaFe$_{2}$(As$_{1-x}$P$_{x}$)$_{2}$, the tuning parameter dependence of structural parameters such as lattice constants are different between on P substitution\cite{S.Kasahara_PRB_2010} and under pressure\cite{R.Mittal_PRB_2011}.
Therefore, the effect of these parameters for superconductivity and AFM fluctuations might be different, although both parameters induce superconductivity\cite{S.Kasahara_PRB_2010,P.L.Alireza_JPCM_2009}.
To date, orbital fluctuations have also been considered to play an important role for the pairing interaction in iron-pnictide superconductors\cite{H.Kontani_PRL_2010}, and in general, it is difficult to measure one of these fluctuations separately.
For this purpose, we would like to point out that $^{31}$P-NMR is one of the best techniques to probe the AFM fluctuations solely, because the nuclear spin of $^{31}$P is 1/2 and electric coupling with the lattice is entirely absent.

In this study, we performed $^{31}$P-NMR measurements on single-crystal BaFe$_{2}$(As$_{1-x}$P$_{x}$)$_{2}$ ($x = 0.22$ and 0.28) under pressure to investigate the effect of pressure on the magnetic properties and the phase diagram. 
At $x = 0.22$, the structural phase transition temperatures $T_{\rm S}$ = 72~K and $T_{\rm N}$ = 55~K are little changed by increasing the pressure up to 2.4~GPa, whereas $T_{\rm c}$ is increased to twice the original value.
On the other hand, for $x = 0.28$, $T_{\rm S}$ = 55~K is quickly suppressed by pressure and $T_{\rm c}$ decreases gradually with increasing pressure.
From a nuclear relaxation rate analysis, we find that the dependence of $T_{\rm c}$ on the Weiss temperature $\theta$ can be quantitatively scaled between pressure and P-content variations, indicating the universal relationship between low-energy AFM fluctuations and superconductivity in BaFe$_{2}$(As$_{1-x}$P$_{x}$)$_{2}$.

Single crystals of BaFe$_{2}$(As$_{1-x}$P$_{x}$)$_{2}$ were prepared as described elsewhere\cite{S.Kasahara_PRB_2010}.
$T_{\rm c}$ = 14.1~K for $x = 0.22$ and 29.1~K for $x = 0.28$ were determined by ac susceptibility measurements using an NMR coil.
Pressure was generated in a piston cylinder-type pressure cell with Daphne 7373 for the $x = 0.22$ samples, and an indenter-type pressure cell with Daphne 7474 for the $x = 0.28$ samples\cite{T.C.Kobayashi_RSI_2007,K.Murata_RSI_2008}.
The applied pressure $P$ was determined from $T_{\rm c}$ of the lead manometer by using the relation of $P$ (GPa) $= [T_{\rm c}(0) - T_{\rm c}(P)]$ (K)/0.364(K/GPa)\cite{A.Eiling_JPFMP_1981,B.Bireckoven_JPESI_1988}.
The $^{31}$P (nuclear spin $I$ = 1/2, nuclear gyromagnetic ratio $^{31}\gamma_{\rm N}/2\pi$ = 17.237~MHz/T, and natural
abundance 100 \%) nuclear spin-lattice relaxation rate $1/T_1$ was determined by fitting the time variation of the spin-echo intensity after the saturation of the nuclear magnetization to a single exponential function across the entire temperature range as shown in Fig.~S1\cite{sup}.

\begin{figure}[!tb]
\vspace*{10pt}
\begin{center}
\includegraphics[width=8.5cm,clip]{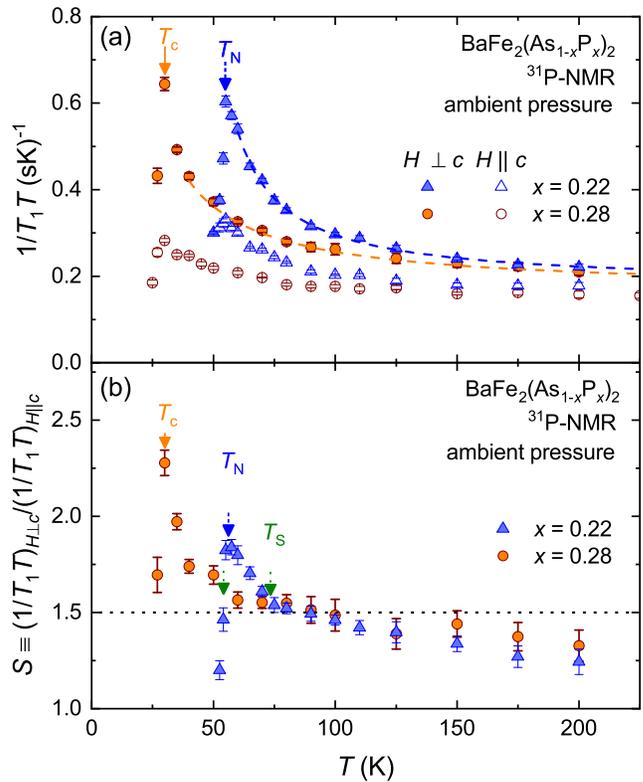}
\end{center}
\caption{(Color online) (a) Temperature dependence of $1/T_1T$ for $H \parallel c$ and $H \perp c$, and (b) the ratio of $1/T_1T$, $S \equiv (1/T_1T)_{H_{\perp c}}/(1/T_1T)_{H_{\parallel c}}$ (b) at ambient pressure measured for the $x = 0.22$ and $x = 0.28$ samples.
The dashed lines of (a) are fitting curves by the Curie-Weiss formula.
The dotted line of (b) indicates the value of 1.5.
}
\label{Fig.2}
\end{figure}
\begin{figure}[!b]
\vspace*{10pt}
\begin{center}
\includegraphics[width=8cm,clip]{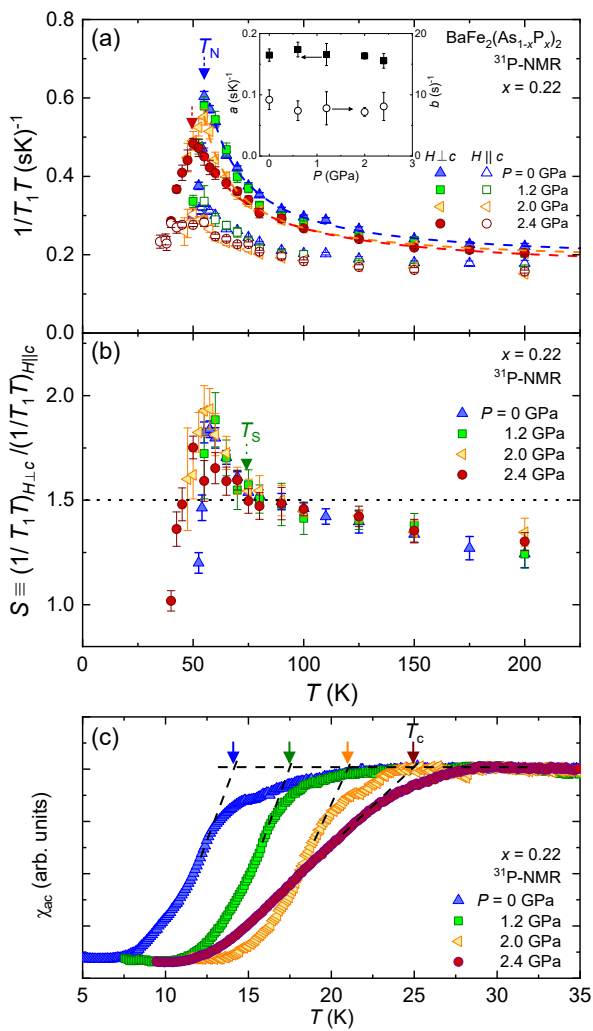}
\end{center}
\caption{(Color online) (a) Temperature dependence of $1/T_1T$ for $H \parallel c$ and $H \perp c$, (b) the ratio of $1/T_1T$, and (c) ac susceptibility for $x = 0.22$ under pressure.
The dashed lines of (a) are fitting curves by the Curie-Weiss formula.
The dotted line of (b) indicates the value of 1.5.
The dashed lines of (c) are intended to guide the eye.
}
\label{Fig.3}
\end{figure}

Figure~\ref{Fig.2}(a) shows the temperature dependence of $1/T_1T$ for $H \parallel c$ and $H \perp c$, and Fig. \ref{Fig.2}(b) the ratio of $1/T_1T$ anisotropy, $S \equiv (1/T_1T)_{H\perp c}/(1/T_1T)_{H\parallel c}$ at $x = 0.22$ and $x = 0.28$ at ambient pressure.
As a result of strong AFM fluctuations, $1/T_1T$ is enhanced toward $T_{\rm N}$ and $T_{\rm c}$ with decreasing $T$; $1/T_1T$ shows a peak at $T_{\rm N}$ by critically slowing down in the $x = 0.22$ sample, and $1/T_1T$ decreases below $T_{\rm c}$ due to the opening of the superconducting gap in the $x = 0.28$ sample.
Below $T_{\rm N}$ and $T_{\rm c}$, the intensity of the NMR signal of the two samples weakens to an extent that $1/T_1T$ could not be measured accurately.
The temperature dependence of $1/T_1T$ for $H \perp c$ is consistent with the previous report measured in the mosaic of single crystals\cite{Y.Nakai_PRL_2010}.
The anisotropy ratio $S$ of $1/T_1T$ is $\sim 1.25 $ at high temperatures in both samples, which originates from the stripe-type spin correlations.
As reported previously\cite{K.Kitagawa_JPSJ_2008,S.Kitagawa_PRB_2010,Y.Nakai_PRB_2012}, the anisotropy ratio of $1/T_1T$ in the system dominated by stripe correlations can be written as,
\begin{align}
S \equiv \frac{(1/T_1T)_{H\perp c}}{(1/T_1T)_{H\parallel c}} = \left|\frac{S_a(\omega_{\rm res})}{S_c(\omega_{\rm res})}\right|^2 + \frac{1}{2},
\end{align}
where $(1/T_1T)_{H\perp c} = \frac{(1/T_1T)_{H\parallel a}+(1/T_1T)_{H\parallel b}}{2}$ and $S_i(\omega)$ ($i = a$ and $c$) denotes the spin fluctuations along the $i$ axis probed by NMR frequency $\omega_{\rm res}$.
Therefore, $S$ becomes 1.5 if the Fe spin fluctuations are isotropic ($|S_{a}|$ = $|S_{c}|$) with the stripe correlations, whereas the ratio becomes higher than 1.5 if in-plane stripe fluctuations develop ($|S_{a}| > |S_{c}|$).
In various iron-based superconductors, a ratio of $\sim 1.5$, suggesting the presence of a stripe AFM correlation, has been observed just above $T_{\rm S}$ or $T_{\rm c}$\cite{K.Kitagawa_JPSJ_2008,S.Kitagawa_PRB_2010,Y.Nakai_PRB_2012,M.Hirano_JPSJ_2012,Z.T.Zhang_PRB_2018}.
Note that $S$ is smaller than 1.5 at high temperatures, originating from the existence of a paramagnetic contribution.
On cooling, $S$ increases more rapidly below $\sim$ $T_{\rm S}$, indicating that the in-plane Fe spin fluctuations increase below the structural phase transition.
The breaking of in-plane four-fold symmetry enhances the stripe-type AFM correlations, because the direction of the AFM correlations is determined. 
In fact, the same enhancement of $S$ below $T_{\rm S}$ was clearly observed in LaFeAs(O$_{1-x}$F$_{x}$)\cite{Y.Nakai_PRB_2012}.
We defined the structural-transition temperature $T_{\rm S}$ as the onset of the increase in $S$ and determined the AFM ordering temperature $T_{\rm N}$ as the peak of $1/T_1T$.

\begin{figure}[!tb]
\vspace*{10pt}
\begin{center}
\includegraphics[width=8cm,clip]{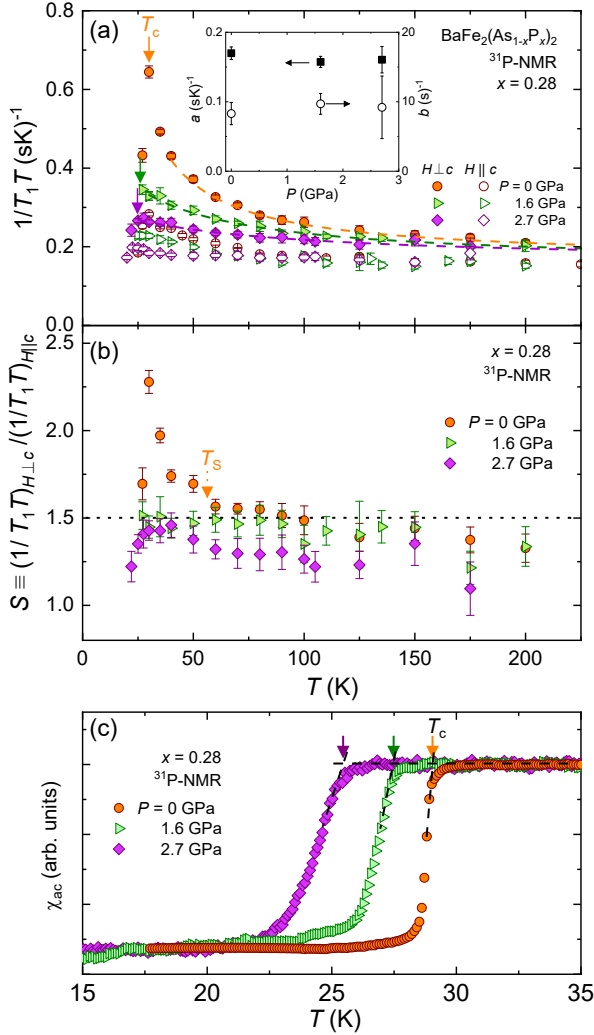}
\end{center}
\caption{(Color online) (a) Temperature dependence of $1/T_1T$ for $H \parallel c$ and $H \perp c$, (b) the ratio of $1/T_1T$, and (c) ac susceptibility at $x = 0.28$ under pressure.
The dashed lines of (a) are fitting curves by the Curie-Weiss formula.
The dotted line of (b) indicates the value of 1.5.
The dashed lines of (c) are intended to guide the eye.
}
\label{Fig.4}
\end{figure}
\begin{figure}[!tb]
\vspace*{10pt}
\begin{center}
\includegraphics[width=8.5cm,clip]{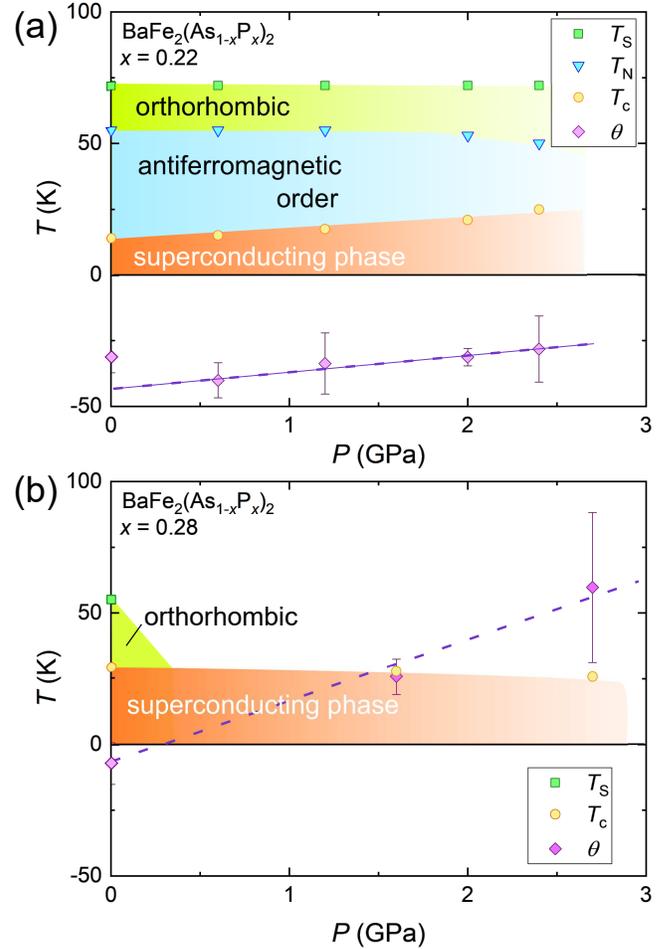}
\end{center}
\caption{(Color online)$P$ - $T$ phase diagram at (a) $x = 0.22$ and (b) $x = 0.28$.
The dashed lines are provided to guide the eye. 
}
\label{Fig.5}
\end{figure}
\begin{figure}[!tb]
\vspace*{10pt}
\begin{center}
\includegraphics[width=8.5cm,clip]{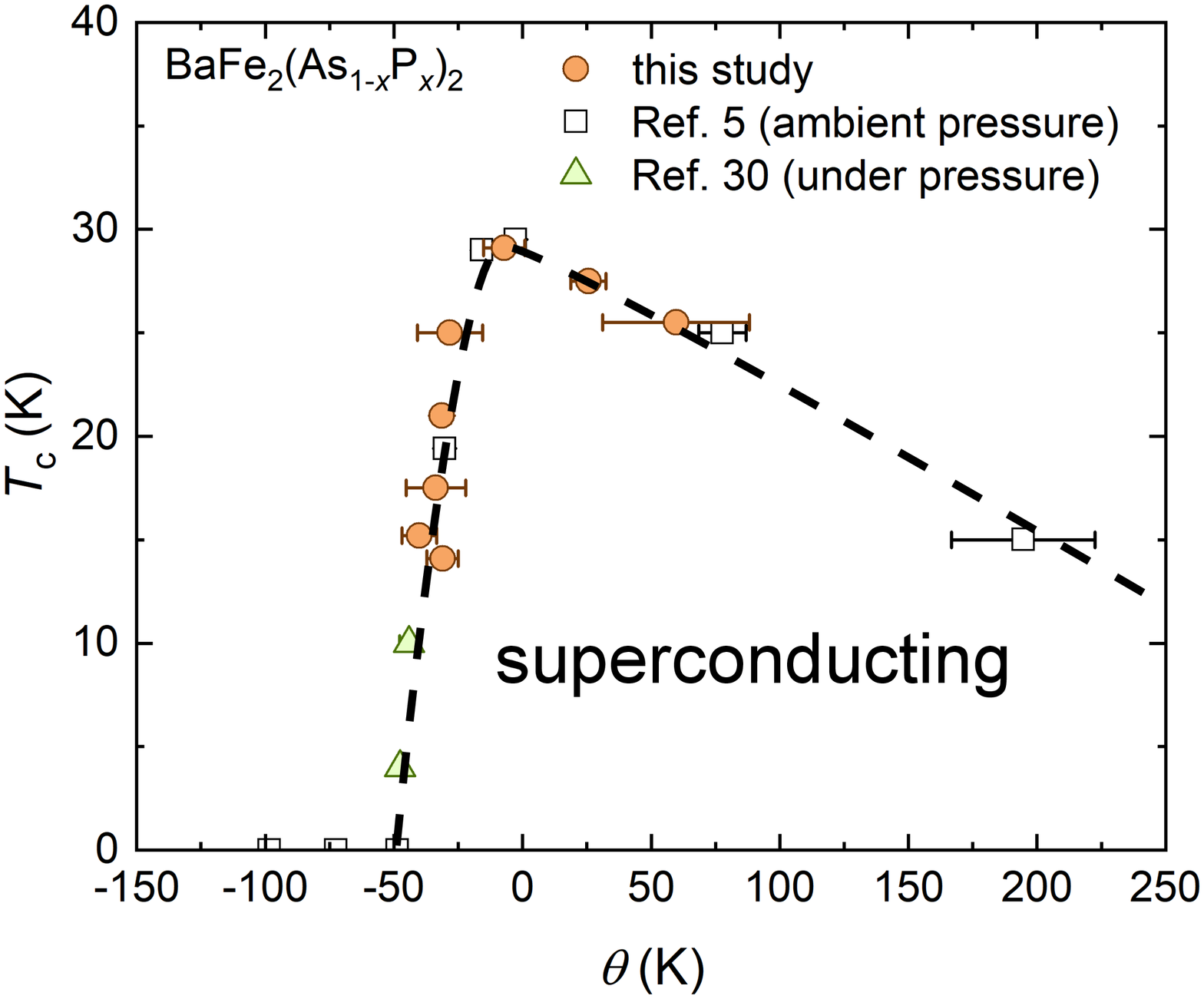}
\end{center}
\caption{(Color online) $\theta$ dependence of $T_{\rm c}$ on BaFe$_{2}$(As$_{1-x}$P$_{x}$)$_{2}$ at ambient pressure\cite{Y.Nakai_PRL_2010} and under pressure.
The dashed curve is added to guide the eye. 
}
\label{Fig.6}
\end{figure}

The pressure dependence of $T_{\rm N}$, $T_{\rm S}$ and $T_{\rm c}$ was investigated with $^{31}$P-NMR and ac susceptibility measurements. 
Figure~\ref{Fig.3} (a) shows the temperature dependence of $1/T_1T$ for $H \parallel c$ and $H \perp c$, Fig.~\ref{Fig.3} (b)  $S$, and Fig.~\ref{Fig.3} (c) the ac susceptibility at $x = 0.22$ under pressure.
Although $T_{\rm c}$ increases from 14.1~K at ambient pressure to 25.0~K at 2.4~GPa, $T_{\rm N}$ and $T_{\rm S}$ show only little changes by pressure.
The limitations of the pressure cell prevented us from reaching the maximum of $T_c$.
In contrast, $1/T_1T$ of the $x = 0.28$ sample is strongly affected by pressure as shown in Fig.~\ref{Fig.4}.
The AFM fluctuations and $T_{\rm S}$ are significantly suppressed by pressure.

To estimate the pressure evolution of the Weiss temperature $\theta$, the temperature dependence of $1/T_1T$ was fitted to Eq.\eqref{eq.2}.
We used the data for $H \perp c$ to compare with the previous results because $1/T_1T$ for $H \perp c$ is determined with the in-plane AFM fluctuations\cite{K.Kitagawa_JPSJ_2008}. 
The fitting parameters of $a$ and $b$ are hardly changed by pressure as shown in the insets of Figs.~\ref{Fig.3} and \ref{Fig.4}.
This indicates that the pressure does not change the density of states, which is consistent with the band-structure calculation and the AFM-fluctuation component significantly\cite{Y.Nakai_PRL_2010}.
We constructed the $P$-$T$ phase diagrams of the $x = 0.22$ and 0.28 samples as shown in Fig.~\ref{Fig.5}.
In the $x = 0.22$ sample, although $T_{\rm N}$, $T_{\rm S}$, and $\theta$ undergo small changes, the increase in $T_{\rm c}$ is large.
In contrast, in the $x = 0.28$ sample, $T_{\rm c}$ gradually decreases with increasing pressure, although $T_{\rm S}$ is abruptly suppressed by pressure and $\theta$ largely increases from $-7$~K at ambient pressure to 60~K at 2.7~GPa, passing through the AFM QCP. 
To understand the relationship between the AFM critical fluctuations and superconductivity, $T_{\rm c}$ is plotted against $\theta$ obtained for both the P-substitution and pressure studies as shown in Fig.~\ref{Fig.6}, where the previous results obtained in the mosaic single crystals of the $x$ = 0.20 sample under pressure\cite{T.Iye_JPSJ_2012} were also analyzed by the same procedure for comparison.      
The dependence of $T_{\rm c}$ on magnetic fluctuations seems asymmetric before and after the AFM QCP, and this asymmetric behavior of $T_{\rm c}$ can be understood in terms of the presence of the AFM phase in the negative $\theta$ region, where the Fermi surfaces partially contribute to the AFM ordering. 
The dependence of $T_{\rm c}$ on $\theta$ obtained by tuning these two parameters is precisely consistent with each other, and this result strongly suggests the existence of a universal relationship between the low-energy AFM fluctuations and superconductivity in BaFe$_{2}$(As$_{1-x}$P$_{x}$)$_{2}$.
Furthermore, we comment on the effect of the nematic fluctuations revealed by measuring $1/T_1$ of $^{75}$As with nuclear quadrupole moment\cite{A.P.Dioguardi_PRL_2016}.
The nematic fluctuations were shown to be enhanced below approximately $T_{\rm S}$ and to possess inhomogeneous glassy dynamics\cite{A.P.Dioguardi_PRL_2016}.
As already mentioned, $1/T_1$ of the $^{31}$P-NMR does not couple with the electric fluctuations related with the lattice dynamics, but only couple with magnetic fluctuations. 
In addition, the deviation from the Curie-Weiss behavior was observed even in $1/T_1T$ of $^{31}$P below $T_{\rm S}$, but the value of $\theta$ was evaluated from the temperature range above $T_{\rm S}$, where the spin fluctuations are homogeneous. Thus, the $\theta$ we evaluated is related to the AFM fluctuations, which are not affected by the nematic fluctuations.  

It is noteworthy that the phase diagram of BaFe$_{2}$(As$_{1-x}$P$_{x}$)$_{2}$ is well summarized by $\theta$ and that the $T_{\rm c}$ maximum is observed near the AFM QCP even when the spin fluctuations are changed by pressure, indicating that the low-temperature properties are determined with the low-energy AFM fluctuations in the normal state, and that the maximum $T_{\rm c}$ near the AFM QCP is not accidental but an intrinsic property.  
Because the application of pressure introduces negligible disorder into the Fe plane and hardly changes the carrier content, and isovalent P substitution shows less significant disorder effects than that of Co or K substitution in BaFe$_2$As$_2$, adjusting both of these parameters is an ideal way to change the strength of electron correlations.   
A simliar $\theta$ dependence of $T_{\rm c}$ was observed in various iron-based superconductors, although maximum $T_{\rm c}$ and the detailed  $\theta$ dependence of $T_{\rm c}$ depend on the system\cite{sup2}.

In conclusion, we performed $^{31}$P-NMR measurements on BaFe$_{2}$(As$_{1-x}$P$_{x}$)$_{2}$ ($x$ = 0.22 and 0.28) under pressure to investigate the relationship between low-energy AFM fluctuations and superconductivity. 
The pressure dependences of $T_{\rm S}$, $T_{\rm N}$, and $T_{\rm c}$ in these two samples are almost the same as the dependences of these temperatures of BaFe$_{2}$(As$_{1-x}$P$_{x}$)$_{2}$ on $x$ at ambient pressure. This indicates the presence of a universal relationship between low-energy AFM fluctuations and superconductivity, with the AFM fluctuation being the key parameter in the case of BaFe$_{2}$(As$_{1-x}$P$_{x}$)$_{2}$.

\section*{Acknowledgments}
The authors acknowledge S. Yonezawa, Y. Maeno, and H. Ikeda for fruitful discussions. 
This work was partially supported by the Kyoto Univ. LTM Center and Grant-in-Aids for Scientific Research (KAKENHI) (Grant Numbers JP15H05882, JP15H05884,  JP15K21732, JP15H05745, JP17K14339, JP19K14657, JP19K04696, and JP19H05824). 


\begin{thebibliography}{32}%
\makeatletter
\providecommand \@ifxundefined [1]{%
 \@ifx{#1\undefined}
}%
\providecommand \@ifnum [1]{%
 \ifnum #1\expandafter \@firstoftwo
 \else \expandafter \@secondoftwo
 \fi
}%
\providecommand \@ifx [1]{%
 \ifx #1\expandafter \@firstoftwo
 \else \expandafter \@secondoftwo
 \fi
}%
\providecommand \natexlab [1]{#1}%
\providecommand \enquote  [1]{``#1''}%
\providecommand \bibnamefont  [1]{#1}%
\providecommand \bibfnamefont [1]{#1}%
\providecommand \citenamefont [1]{#1}%
\providecommand \href@noop [0]{\@secondoftwo}%
\providecommand \href [0]{\begingroup \@sanitize@url \@href}%
\providecommand \@href[1]{\@@startlink{#1}\@@href}%
\providecommand \@@href[1]{\endgroup#1\@@endlink}%
\providecommand \@sanitize@url [0]{\catcode `\\12\catcode `\$12\catcode
  `\&12\catcode `\#12\catcode `\^12\catcode `\_12\catcode `\%12\relax}%
\providecommand \@@startlink[1]{}%
\providecommand \@@endlink[0]{}%
\providecommand \url  [0]{\begingroup\@sanitize@url \@url }%
\providecommand \@url [1]{\endgroup\@href {#1}{\urlprefix }}%
\providecommand \urlprefix  [0]{URL }%
\providecommand \Eprint [0]{\href }%
\providecommand \doibase [0]{http://dx.doi.org/}%
\providecommand \selectlanguage [0]{\@gobble}%
\providecommand \bibinfo  [0]{\@secondoftwo}%
\providecommand \bibfield  [0]{\@secondoftwo}%
\providecommand \translation [1]{[#1]}%
\providecommand \BibitemOpen [0]{}%
\providecommand \bibitemStop [0]{}%
\providecommand \bibitemNoStop [0]{.\EOS\space}%
\providecommand \EOS [0]{\spacefactor3000\relax}%
\providecommand \BibitemShut  [1]{\csname bibitem#1\endcsname}%
\let\auto@bib@innerbib\@empty
\bibitem [{\citenamefont {Bardeen}\ \emph {et~al.}(1957)\citenamefont
  {Bardeen}, \citenamefont {Cooper},\ and\ \citenamefont
  {Schrieffer}}]{J.Bardeen_PR_1957}%
  \BibitemOpen
  \bibfield  {author} {\bibinfo {author} {\bibfnamefont {J.}~\bibnamefont
  {Bardeen}}, \bibinfo {author} {\bibfnamefont {L.~N.}\ \bibnamefont {Cooper}},
  \ and\ \bibinfo {author} {\bibfnamefont {J.~R.}\ \bibnamefont {Schrieffer}},\
  }\href@noop {} {\bibfield  {journal} {\bibinfo  {journal} {Phys. Rev.}\
  }\textbf {\bibinfo {volume} {108}},\ \bibinfo {pages} {1175} (\bibinfo {year}
  {1957})}\BibitemShut {NoStop}%
\bibitem [{\citenamefont {Moriya}\ and\ \citenamefont
  {Ueda}(1994)}]{T.Moriya_JPSJ_1994}%
  \BibitemOpen
  \bibfield  {author} {\bibinfo {author} {\bibfnamefont {T.}~\bibnamefont
  {Moriya}}\ and\ \bibinfo {author} {\bibfnamefont {K.}~\bibnamefont {Ueda}},\
  }\href@noop {} {\bibfield  {journal} {\bibinfo  {journal} {J. Phys. Soc.
  Jpn.}\ }\textbf {\bibinfo {volume} {63}},\ \bibinfo {pages} {1871} (\bibinfo
  {year} {1994})}\BibitemShut {NoStop}%
\bibitem [{\citenamefont {Nakai}\ \emph {et~al.}(2013)\citenamefont {Nakai},
  \citenamefont {Iye}, \citenamefont {Kitagawa}, \citenamefont {Ishida},
  \citenamefont {Kasahara}, \citenamefont {Shibauchi}, \citenamefont {Matsuda},
  \citenamefont {Ikeda},\ and\ \citenamefont {Terashima}}]{Y.Nakai_PRB_2013}%
  \BibitemOpen
  \bibfield  {author} {\bibinfo {author} {\bibfnamefont {Y.}~\bibnamefont
  {Nakai}}, \bibinfo {author} {\bibfnamefont {T.}~\bibnamefont {Iye}}, \bibinfo
  {author} {\bibfnamefont {S.}~\bibnamefont {Kitagawa}}, \bibinfo {author}
  {\bibfnamefont {K.}~\bibnamefont {Ishida}}, \bibinfo {author} {\bibfnamefont
  {S.}~\bibnamefont {Kasahara}}, \bibinfo {author} {\bibfnamefont
  {T.}~\bibnamefont {Shibauchi}}, \bibinfo {author} {\bibfnamefont
  {Y.}~\bibnamefont {Matsuda}}, \bibinfo {author} {\bibfnamefont
  {H.}~\bibnamefont {Ikeda}}, \ and\ \bibinfo {author} {\bibfnamefont
  {T.}~\bibnamefont {Terashima}},\ }\href@noop {} {\bibfield  {journal}
  {\bibinfo  {journal} {Phys. Rev. B}\ }\textbf {\bibinfo {volume} {87}},\
  \bibinfo {pages} {174507} (\bibinfo {year} {2013})}\BibitemShut {NoStop}%
\bibitem [{\citenamefont {Sarrao}\ \emph {et~al.}(2015)\citenamefont {Sarrao},
  \citenamefont {Bauer}, \citenamefont {Mitchell}, \citenamefont {Tobash},\
  and\ \citenamefont {Thompson}}]{J.L.Sarrao_PhysicaC_2015}%
  \BibitemOpen
  \bibfield  {author} {\bibinfo {author} {\bibfnamefont {J.~L.}\ \bibnamefont
  {Sarrao}}, \bibinfo {author} {\bibfnamefont {E.~D.}\ \bibnamefont {Bauer}},
  \bibinfo {author} {\bibfnamefont {J.~N.}\ \bibnamefont {Mitchell}}, \bibinfo
  {author} {\bibfnamefont {P.~H.}\ \bibnamefont {Tobash}}, \ and\ \bibinfo
  {author} {\bibfnamefont {J.~D.}\ \bibnamefont {Thompson}},\ }\href@noop {}
  {\bibfield  {journal} {\bibinfo  {journal} {Physica C}\ }\textbf {\bibinfo
  {volume} {514}},\ \bibinfo {pages} {184} (\bibinfo {year}
  {2015})}\BibitemShut {NoStop}%
\bibitem [{\citenamefont {Nakai}\ \emph {et~al.}(2010)\citenamefont {Nakai},
  \citenamefont {Iye}, \citenamefont {Kitagawa}, \citenamefont {Ishida},
  \citenamefont {Ikeda}, \citenamefont {Kasahara}, \citenamefont {Shishido},
  \citenamefont {Shibauchi}, \citenamefont {Matsuda},\ and\ \citenamefont
  {Terashima}}]{Y.Nakai_PRL_2010}%
  \BibitemOpen
  \bibfield  {author} {\bibinfo {author} {\bibfnamefont {Y.}~\bibnamefont
  {Nakai}}, \bibinfo {author} {\bibfnamefont {T.}~\bibnamefont {Iye}}, \bibinfo
  {author} {\bibfnamefont {S.}~\bibnamefont {Kitagawa}}, \bibinfo {author}
  {\bibfnamefont {K.}~\bibnamefont {Ishida}}, \bibinfo {author} {\bibfnamefont
  {H.}~\bibnamefont {Ikeda}}, \bibinfo {author} {\bibfnamefont
  {S.}~\bibnamefont {Kasahara}}, \bibinfo {author} {\bibfnamefont
  {H.}~\bibnamefont {Shishido}}, \bibinfo {author} {\bibfnamefont
  {T.}~\bibnamefont {Shibauchi}}, \bibinfo {author} {\bibfnamefont
  {Y.}~\bibnamefont {Matsuda}}, \ and\ \bibinfo {author} {\bibfnamefont
  {T.}~\bibnamefont {Terashima}},\ }\href@noop {} {\bibfield  {journal}
  {\bibinfo  {journal} {Phys. Rev. Lett.}\ }\textbf {\bibinfo {volume} {105}},\
  \bibinfo {pages} {107003} (\bibinfo {year} {2010})}\BibitemShut {NoStop}%
\bibitem [{\citenamefont {Hashimoto}\ \emph {et~al.}(2012)\citenamefont
  {Hashimoto}, \citenamefont {Cho}, \citenamefont {Shibauchi}, \citenamefont
  {Kasahara}, \citenamefont {Mizukami}, \citenamefont {Katsumata},
  \citenamefont {Tsuruhara}, \citenamefont {Terashima}, \citenamefont {Ikeda},
  \citenamefont {Tanatar}, \citenamefont {Kitano}, \citenamefont {Salovich},
  \citenamefont {Giannetta}, \citenamefont {Walmsley}, \citenamefont
  {Carrington}, \citenamefont {Prozorov},\ and\ \citenamefont
  {Matsuda}}]{K.Hashimoto_Science_2012}%
  \BibitemOpen
  \bibfield  {author} {\bibinfo {author} {\bibfnamefont {K.}~\bibnamefont
  {Hashimoto}}, \bibinfo {author} {\bibfnamefont {K.}~\bibnamefont {Cho}},
  \bibinfo {author} {\bibfnamefont {T.}~\bibnamefont {Shibauchi}}, \bibinfo
  {author} {\bibfnamefont {S.}~\bibnamefont {Kasahara}}, \bibinfo {author}
  {\bibfnamefont {Y.}~\bibnamefont {Mizukami}}, \bibinfo {author}
  {\bibfnamefont {R.}~\bibnamefont {Katsumata}}, \bibinfo {author}
  {\bibfnamefont {Y.}~\bibnamefont {Tsuruhara}}, \bibinfo {author}
  {\bibfnamefont {T.}~\bibnamefont {Terashima}}, \bibinfo {author}
  {\bibfnamefont {H.}~\bibnamefont {Ikeda}}, \bibinfo {author} {\bibfnamefont
  {M.~A.}\ \bibnamefont {Tanatar}}, \bibinfo {author} {\bibfnamefont
  {H.}~\bibnamefont {Kitano}}, \bibinfo {author} {\bibfnamefont
  {N.}~\bibnamefont {Salovich}}, \bibinfo {author} {\bibfnamefont {R.~W.}\
  \bibnamefont {Giannetta}}, \bibinfo {author} {\bibfnamefont {P.}~\bibnamefont
  {Walmsley}}, \bibinfo {author} {\bibfnamefont {A.}~\bibnamefont
  {Carrington}}, \bibinfo {author} {\bibfnamefont {R.}~\bibnamefont
  {Prozorov}}, \ and\ \bibinfo {author} {\bibfnamefont {Y.}~\bibnamefont
  {Matsuda}},\ }\href@noop {} {\bibfield  {journal} {\bibinfo  {journal}
  {Science}\ }\textbf {\bibinfo {volume} {336}},\ \bibinfo {pages} {1554}
  (\bibinfo {year} {2012})}\BibitemShut {NoStop}%
\bibitem [{\citenamefont {Moriya}\ \emph {et~al.}(1990)\citenamefont {Moriya},
  \citenamefont {Takahashi},\ and\ \citenamefont {Ueda}}]{T.Moriya_JPSJ_1990}%
  \BibitemOpen
  \bibfield  {author} {\bibinfo {author} {\bibfnamefont {T.}~\bibnamefont
  {Moriya}}, \bibinfo {author} {\bibfnamefont {Y.}~\bibnamefont {Takahashi}}, \
  and\ \bibinfo {author} {\bibfnamefont {K.}~\bibnamefont {Ueda}},\ }\href@noop
  {} {\bibfield  {journal} {\bibinfo  {journal} {J. Phys. Soc. Jpn.}\ }\textbf
  {\bibinfo {volume} {59}},\ \bibinfo {pages} {2905} (\bibinfo {year}
  {1990})}\BibitemShut {NoStop}%
\bibitem [{\citenamefont {Ning}\ \emph {et~al.}(2010)\citenamefont {Ning},
  \citenamefont {Ahilan}, \citenamefont {Imai}, \citenamefont {Sefat},
  \citenamefont {McGuire}, \citenamefont {Sales}, \citenamefont {Mandrus},
  \citenamefont {Cheng}, \citenamefont {Shen},\ and\ \citenamefont
  {Wen}}]{F.L.Ning_PRL_2010}%
  \BibitemOpen
  \bibfield  {author} {\bibinfo {author} {\bibfnamefont {F.~L.}\ \bibnamefont
  {Ning}}, \bibinfo {author} {\bibfnamefont {K.}~\bibnamefont {Ahilan}},
  \bibinfo {author} {\bibfnamefont {T.}~\bibnamefont {Imai}}, \bibinfo {author}
  {\bibfnamefont {A.~S.}\ \bibnamefont {Sefat}}, \bibinfo {author}
  {\bibfnamefont {M.~A.}\ \bibnamefont {McGuire}}, \bibinfo {author}
  {\bibfnamefont {B.~C.}\ \bibnamefont {Sales}}, \bibinfo {author}
  {\bibfnamefont {D.}~\bibnamefont {Mandrus}}, \bibinfo {author} {\bibfnamefont
  {P.}~\bibnamefont {Cheng}}, \bibinfo {author} {\bibfnamefont
  {B.}~\bibnamefont {Shen}}, \ and\ \bibinfo {author} {\bibfnamefont {H.-H.}\
  \bibnamefont {Wen}},\ }\href@noop {} {\bibfield  {journal} {\bibinfo
  {journal} {Phys. Rev. Lett.}\ }\textbf {\bibinfo {volume} {104}},\ \bibinfo
  {pages} {037001} (\bibinfo {year} {2010})}\BibitemShut {NoStop}%
\bibitem [{\citenamefont {Zhou}\ \emph {et~al.}(2013)\citenamefont {Zhou},
  \citenamefont {Li}, \citenamefont {Yang}, \citenamefont {Sun}, \citenamefont
  {Lin},\ and\ \citenamefont {qing Zheng}}]{R.Zhou_NatCommun_2013}%
  \BibitemOpen
  \bibfield  {author} {\bibinfo {author} {\bibfnamefont {R.}~\bibnamefont
  {Zhou}}, \bibinfo {author} {\bibfnamefont {Z.}~\bibnamefont {Li}}, \bibinfo
  {author} {\bibfnamefont {J.}~\bibnamefont {Yang}}, \bibinfo {author}
  {\bibfnamefont {D.}~\bibnamefont {Sun}}, \bibinfo {author} {\bibfnamefont
  {C.}~\bibnamefont {Lin}}, \ and\ \bibinfo {author} {\bibfnamefont
  {G.}~\bibnamefont {qing Zheng}},\ }\href@noop {} {\bibfield  {journal}
  {\bibinfo  {journal} {Nat. Commun.}\ }\textbf {\bibinfo {volume} {4}},\
  \bibinfo {pages} {2265} (\bibinfo {year} {2013})}\BibitemShut {NoStop}%
\bibitem [{\citenamefont {Ji}\ \emph {et~al.}(2013)\citenamefont {Ji},
  \citenamefont {Zhang}, \citenamefont {Ma}, \citenamefont {Fan}, \citenamefont
  {Wang}, \citenamefont {Dai}, \citenamefont {Tan}, \citenamefont {Song},
  \citenamefont {Zhang}, \citenamefont {Dai}, \citenamefont {Normand},\ and\
  \citenamefont {Yu}}]{G.F.Ji_PRL_2013}%
  \BibitemOpen
  \bibfield  {author} {\bibinfo {author} {\bibfnamefont {G.~F.}\ \bibnamefont
  {Ji}}, \bibinfo {author} {\bibfnamefont {J.~S.}\ \bibnamefont {Zhang}},
  \bibinfo {author} {\bibfnamefont {L.}~\bibnamefont {Ma}}, \bibinfo {author}
  {\bibfnamefont {P.}~\bibnamefont {Fan}}, \bibinfo {author} {\bibfnamefont
  {P.~S.}\ \bibnamefont {Wang}}, \bibinfo {author} {\bibfnamefont
  {J.}~\bibnamefont {Dai}}, \bibinfo {author} {\bibfnamefont {G.~T.}\
  \bibnamefont {Tan}}, \bibinfo {author} {\bibfnamefont {Y.}~\bibnamefont
  {Song}}, \bibinfo {author} {\bibfnamefont {C.~L.}\ \bibnamefont {Zhang}},
  \bibinfo {author} {\bibfnamefont {P.}~\bibnamefont {Dai}}, \bibinfo {author}
  {\bibfnamefont {B.}~\bibnamefont {Normand}}, \ and\ \bibinfo {author}
  {\bibfnamefont {W.}~\bibnamefont {Yu}},\ }\href@noop {} {\bibfield  {journal}
  {\bibinfo  {journal} {Phys. Rev. Lett.}\ }\textbf {\bibinfo {volume} {111}},\
  \bibinfo {pages} {107004} (\bibinfo {year} {2013})}\BibitemShut {NoStop}%
\bibitem [{\citenamefont {Miyamoto}\ \emph {et~al.}(2015)\citenamefont
  {Miyamoto}, \citenamefont {Mukuda}, \citenamefont {Kobayashi}, \citenamefont
  {Yashima}, \citenamefont {Kitaoka}, \citenamefont {Miyasaka},\ and\
  \citenamefont {Tajima}}]{M.Miyamoto_PRB_2015}%
  \BibitemOpen
  \bibfield  {author} {\bibinfo {author} {\bibfnamefont {M.}~\bibnamefont
  {Miyamoto}}, \bibinfo {author} {\bibfnamefont {H.}~\bibnamefont {Mukuda}},
  \bibinfo {author} {\bibfnamefont {T.}~\bibnamefont {Kobayashi}}, \bibinfo
  {author} {\bibfnamefont {M.}~\bibnamefont {Yashima}}, \bibinfo {author}
  {\bibfnamefont {Y.}~\bibnamefont {Kitaoka}}, \bibinfo {author} {\bibfnamefont
  {S.}~\bibnamefont {Miyasaka}}, \ and\ \bibinfo {author} {\bibfnamefont
  {S.}~\bibnamefont {Tajima}},\ }\href@noop {} {\bibfield  {journal} {\bibinfo
  {journal} {Phys. Rev. B}\ }\textbf {\bibinfo {volume} {92}},\ \bibinfo
  {pages} {125154} (\bibinfo {year} {2015})}\BibitemShut {NoStop}%
\bibitem [{\citenamefont {E.~Klintberg}\ \emph {et~al.}(2010)\citenamefont
  {E.~Klintberg}, \citenamefont {K.~Goh}, \citenamefont {Kasahara},
  \citenamefont {Nakai}, \citenamefont {Ishida}, \citenamefont {Sutherland},
  \citenamefont {Shibauchi}, \citenamefont {Matsuda},\ and\ \citenamefont
  {Terashima}}]{E.Klintberg_JPSJ_2010}%
  \BibitemOpen
  \bibfield  {author} {\bibinfo {author} {\bibfnamefont {L.}~\bibnamefont
  {E.~Klintberg}}, \bibinfo {author} {\bibfnamefont {S.}~\bibnamefont
  {K.~Goh}}, \bibinfo {author} {\bibfnamefont {S.}~\bibnamefont {Kasahara}},
  \bibinfo {author} {\bibfnamefont {Y.}~\bibnamefont {Nakai}}, \bibinfo
  {author} {\bibfnamefont {K.}~\bibnamefont {Ishida}}, \bibinfo {author}
  {\bibfnamefont {M.}~\bibnamefont {Sutherland}}, \bibinfo {author}
  {\bibfnamefont {T.}~\bibnamefont {Shibauchi}}, \bibinfo {author}
  {\bibfnamefont {Y.}~\bibnamefont {Matsuda}}, \ and\ \bibinfo {author}
  {\bibfnamefont {T.}~\bibnamefont {Terashima}},\ }\href@noop {} {\bibfield
  {journal} {\bibinfo  {journal} {J. Phys. Soc. Jpn.}\ }\textbf {\bibinfo
  {volume} {79}},\ \bibinfo {pages} {123706} (\bibinfo {year}
  {2010})}\BibitemShut {NoStop}%
\bibitem [{\citenamefont {Hosoi}\ \emph {et~al.}(2016)\citenamefont {Hosoi},
  \citenamefont {Matsuura}, \citenamefont {Ishida}, \citenamefont {Wang},
  \citenamefont {Mizukami}, \citenamefont {Watashige}, \citenamefont
  {Kasahara}, \citenamefont {Matsuda},\ and\ \citenamefont
  {Shibauchi}}]{S.Hosoi_PNAS_2016}%
  \BibitemOpen
  \bibfield  {author} {\bibinfo {author} {\bibfnamefont {S.}~\bibnamefont
  {Hosoi}}, \bibinfo {author} {\bibfnamefont {K.}~\bibnamefont {Matsuura}},
  \bibinfo {author} {\bibfnamefont {K.}~\bibnamefont {Ishida}}, \bibinfo
  {author} {\bibfnamefont {H.}~\bibnamefont {Wang}}, \bibinfo {author}
  {\bibfnamefont {Y.}~\bibnamefont {Mizukami}}, \bibinfo {author}
  {\bibfnamefont {T.}~\bibnamefont {Watashige}}, \bibinfo {author}
  {\bibfnamefont {S.}~\bibnamefont {Kasahara}}, \bibinfo {author}
  {\bibfnamefont {Y.}~\bibnamefont {Matsuda}}, \ and\ \bibinfo {author}
  {\bibfnamefont {T.}~\bibnamefont {Shibauchi}},\ }\href@noop {} {\bibfield
  {journal} {\bibinfo  {journal} {Proc. Natl. Acad. Sci. USA}\ }\textbf
  {\bibinfo {volume} {113}},\ \bibinfo {pages} {8139} (\bibinfo {year}
  {2016})}\BibitemShut {NoStop}%
\bibitem [{\citenamefont {Sun}\ \emph {et~al.}(2016)\citenamefont {Sun},
  \citenamefont {Matsuura}, \citenamefont {Ye}, \citenamefont {Mizukami},
  \citenamefont {Shimozawa}, \citenamefont {Matsubayashi}, \citenamefont
  {Yamashita}, \citenamefont {Watashige}, \citenamefont {Kasahara},
  \citenamefont {Matsuda}, \citenamefont {Yan}, \citenamefont {Sales},
  \citenamefont {Uwatoko}, \citenamefont {Cheng},\ and\ \citenamefont
  {Shibauchi}}]{J.P.Sun_NatCommun_2016}%
  \BibitemOpen
  \bibfield  {author} {\bibinfo {author} {\bibfnamefont {J.~P.}\ \bibnamefont
  {Sun}}, \bibinfo {author} {\bibfnamefont {K.}~\bibnamefont {Matsuura}},
  \bibinfo {author} {\bibfnamefont {G.~Z.}\ \bibnamefont {Ye}}, \bibinfo
  {author} {\bibfnamefont {Y.}~\bibnamefont {Mizukami}}, \bibinfo {author}
  {\bibfnamefont {M.}~\bibnamefont {Shimozawa}}, \bibinfo {author}
  {\bibfnamefont {K.}~\bibnamefont {Matsubayashi}}, \bibinfo {author}
  {\bibfnamefont {M.}~\bibnamefont {Yamashita}}, \bibinfo {author}
  {\bibfnamefont {T.}~\bibnamefont {Watashige}}, \bibinfo {author}
  {\bibfnamefont {S.}~\bibnamefont {Kasahara}}, \bibinfo {author}
  {\bibfnamefont {Y.}~\bibnamefont {Matsuda}}, \bibinfo {author} {\bibfnamefont
  {J.~Q.}\ \bibnamefont {Yan}}, \bibinfo {author} {\bibfnamefont {B.~C.}\
  \bibnamefont {Sales}}, \bibinfo {author} {\bibfnamefont {Y.}~\bibnamefont
  {Uwatoko}}, \bibinfo {author} {\bibfnamefont {J.~G.}\ \bibnamefont {Cheng}},
  \ and\ \bibinfo {author} {\bibfnamefont {T.}~\bibnamefont {Shibauchi}},\
  }\href@noop {} {\bibfield  {journal} {\bibinfo  {journal} {Nat. Commun.}\
  }\textbf {\bibinfo {volume} {7}},\ \bibinfo {pages} {12146} (\bibinfo {year}
  {2016})}\BibitemShut {NoStop}%
\bibitem [{\citenamefont {Kawasaki}\ \emph {et~al.}(2006)\citenamefont
  {Kawasaki}, \citenamefont {Yashima}, \citenamefont {Mugino}, \citenamefont
  {Mukuda}, \citenamefont {Kitaoka}, \citenamefont {Shishido},\ and\
  \citenamefont {\ifmmode~\bar{O}\else \={O}\fi{}nuki}}]{S.Kawasaki_PRL_2006}%
  \BibitemOpen
  \bibfield  {author} {\bibinfo {author} {\bibfnamefont {S.}~\bibnamefont
  {Kawasaki}}, \bibinfo {author} {\bibfnamefont {M.}~\bibnamefont {Yashima}},
  \bibinfo {author} {\bibfnamefont {Y.}~\bibnamefont {Mugino}}, \bibinfo
  {author} {\bibfnamefont {H.}~\bibnamefont {Mukuda}}, \bibinfo {author}
  {\bibfnamefont {Y.}~\bibnamefont {Kitaoka}}, \bibinfo {author} {\bibfnamefont
  {H.}~\bibnamefont {Shishido}}, \ and\ \bibinfo {author} {\bibfnamefont
  {Y.}~\bibnamefont {\ifmmode~\bar{O}\else \={O}\fi{}nuki}},\ }\href@noop {}
  {\bibfield  {journal} {\bibinfo  {journal} {Phys. Rev. Lett.}\ }\textbf
  {\bibinfo {volume} {96}},\ \bibinfo {pages} {147001} (\bibinfo {year}
  {2006})}\BibitemShut {NoStop}%
\bibitem [{\citenamefont {Kasahara}\ \emph {et~al.}(2010)\citenamefont
  {Kasahara}, \citenamefont {Shibauchi}, \citenamefont {Hashimoto},
  \citenamefont {Ikada}, \citenamefont {Tonegawa}, \citenamefont {Okazaki},
  \citenamefont {Shishido}, \citenamefont {Ikeda}, \citenamefont {Takeya},
  \citenamefont {Hirata}, \citenamefont {Terashima},\ and\ \citenamefont
  {Matsuda}}]{S.Kasahara_PRB_2010}%
  \BibitemOpen
  \bibfield  {author} {\bibinfo {author} {\bibfnamefont {S.}~\bibnamefont
  {Kasahara}}, \bibinfo {author} {\bibfnamefont {T.}~\bibnamefont {Shibauchi}},
  \bibinfo {author} {\bibfnamefont {K.}~\bibnamefont {Hashimoto}}, \bibinfo
  {author} {\bibfnamefont {K.}~\bibnamefont {Ikada}}, \bibinfo {author}
  {\bibfnamefont {S.}~\bibnamefont {Tonegawa}}, \bibinfo {author}
  {\bibfnamefont {R.}~\bibnamefont {Okazaki}}, \bibinfo {author} {\bibfnamefont
  {H.}~\bibnamefont {Shishido}}, \bibinfo {author} {\bibfnamefont
  {H.}~\bibnamefont {Ikeda}}, \bibinfo {author} {\bibfnamefont
  {H.}~\bibnamefont {Takeya}}, \bibinfo {author} {\bibfnamefont
  {K.}~\bibnamefont {Hirata}}, \bibinfo {author} {\bibfnamefont
  {T.}~\bibnamefont {Terashima}}, \ and\ \bibinfo {author} {\bibfnamefont
  {Y.}~\bibnamefont {Matsuda}},\ }\href@noop {} {\bibfield  {journal} {\bibinfo
   {journal} {Phys. Rev. B}\ }\textbf {\bibinfo {volume} {81}},\ \bibinfo
  {pages} {184519} (\bibinfo {year} {2010})}\BibitemShut {NoStop}%
\bibitem [{\citenamefont {Mittal}\ \emph {et~al.}(2011)\citenamefont {Mittal},
  \citenamefont {Mishra}, \citenamefont {Chaplot}, \citenamefont {Ovsyannikov},
  \citenamefont {Greenberg}, \citenamefont {Trots}, \citenamefont
  {Dubrovinsky}, \citenamefont {Su}, \citenamefont {Brueckel}, \citenamefont
  {Matsuishi}, \citenamefont {Hosono},\ and\ \citenamefont
  {Garbarino}}]{R.Mittal_PRB_2011}%
  \BibitemOpen
  \bibfield  {author} {\bibinfo {author} {\bibfnamefont {R.}~\bibnamefont
  {Mittal}}, \bibinfo {author} {\bibfnamefont {S.~K.}\ \bibnamefont {Mishra}},
  \bibinfo {author} {\bibfnamefont {S.~L.}\ \bibnamefont {Chaplot}}, \bibinfo
  {author} {\bibfnamefont {S.~V.}\ \bibnamefont {Ovsyannikov}}, \bibinfo
  {author} {\bibfnamefont {E.}~\bibnamefont {Greenberg}}, \bibinfo {author}
  {\bibfnamefont {D.~M.}\ \bibnamefont {Trots}}, \bibinfo {author}
  {\bibfnamefont {L.}~\bibnamefont {Dubrovinsky}}, \bibinfo {author}
  {\bibfnamefont {Y.}~\bibnamefont {Su}}, \bibinfo {author} {\bibfnamefont
  {T.}~\bibnamefont {Brueckel}}, \bibinfo {author} {\bibfnamefont
  {S.}~\bibnamefont {Matsuishi}}, \bibinfo {author} {\bibfnamefont
  {H.}~\bibnamefont {Hosono}}, \ and\ \bibinfo {author} {\bibfnamefont
  {G.}~\bibnamefont {Garbarino}},\ }\href@noop {} {\bibfield  {journal}
  {\bibinfo  {journal} {Phys. Rev. B}\ }\textbf {\bibinfo {volume} {83}},\
  \bibinfo {pages} {054503} (\bibinfo {year} {2011})}\BibitemShut {NoStop}%
\bibitem [{\citenamefont {Alireza}\ \emph {et~al.}(2009)\citenamefont
  {Alireza}, \citenamefont {Ko}, \citenamefont {Gillett}, \citenamefont
  {Petrone}, \citenamefont {Cole}, \citenamefont {Lonzarich},\ and\
  \citenamefont {Sebastian}}]{P.L.Alireza_JPCM_2009}%
  \BibitemOpen
  \bibfield  {author} {\bibinfo {author} {\bibfnamefont {P.~L.}\ \bibnamefont
  {Alireza}}, \bibinfo {author} {\bibfnamefont {Y.~T.~C.}\ \bibnamefont {Ko}},
  \bibinfo {author} {\bibfnamefont {J.}~\bibnamefont {Gillett}}, \bibinfo
  {author} {\bibfnamefont {C.~M.}\ \bibnamefont {Petrone}}, \bibinfo {author}
  {\bibfnamefont {J.~M.}\ \bibnamefont {Cole}}, \bibinfo {author}
  {\bibfnamefont {G.~G.}\ \bibnamefont {Lonzarich}}, \ and\ \bibinfo {author}
  {\bibfnamefont {S.~E.}\ \bibnamefont {Sebastian}},\ }\href@noop {} {\bibfield
   {journal} {\bibinfo  {journal} {J. Phys : Condens. Matter}\ }\textbf
  {\bibinfo {volume} {21}},\ \bibinfo {pages} {012208} (\bibinfo {year}
  {2009})}\BibitemShut {NoStop}%
\bibitem [{\citenamefont {Kontani}\ and\ \citenamefont
  {Onari}(2010)}]{H.Kontani_PRL_2010}%
  \BibitemOpen
  \bibfield  {author} {\bibinfo {author} {\bibfnamefont {H.}~\bibnamefont
  {Kontani}}\ and\ \bibinfo {author} {\bibfnamefont {S.}~\bibnamefont
  {Onari}},\ }\href@noop {} {\bibfield  {journal} {\bibinfo  {journal} {Phys.
  Rev. Lett.}\ }\textbf {\bibinfo {volume} {104}},\ \bibinfo {pages} {157001}
  (\bibinfo {year} {2010})}\BibitemShut {NoStop}%
\bibitem [{\citenamefont {Kobayashi}\ \emph {et~al.}(2007)\citenamefont
  {Kobayashi}, \citenamefont {Hidaka}, \citenamefont {Kotegawa}, \citenamefont
  {Fujiwara},\ and\ \citenamefont {Eremets}}]{T.C.Kobayashi_RSI_2007}%
  \BibitemOpen
  \bibfield  {author} {\bibinfo {author} {\bibfnamefont {T.~C.}\ \bibnamefont
  {Kobayashi}}, \bibinfo {author} {\bibfnamefont {H.}~\bibnamefont {Hidaka}},
  \bibinfo {author} {\bibfnamefont {H.}~\bibnamefont {Kotegawa}}, \bibinfo
  {author} {\bibfnamefont {K.}~\bibnamefont {Fujiwara}}, \ and\ \bibinfo
  {author} {\bibfnamefont {M.~I.}\ \bibnamefont {Eremets}},\ }\href@noop {}
  {\bibfield  {journal} {\bibinfo  {journal} {Rev. Sci. Instrum.}\ }\textbf
  {\bibinfo {volume} {78}},\ \bibinfo {pages} {023909} (\bibinfo {year}
  {2007})}\BibitemShut {NoStop}%
\bibitem [{\citenamefont {Murata}\ \emph {et~al.}(2008)\citenamefont {Murata},
  \citenamefont {Yokogawa}, \citenamefont {Yoshino}, \citenamefont {Klotz},
  \citenamefont {Munsch}, \citenamefont {Irizawa}, \citenamefont {Nishiyama},
  \citenamefont {Iizuka}, \citenamefont {Nanba}, \citenamefont {Okada},
  \citenamefont {Shiraga},\ and\ \citenamefont {Aoyama}}]{K.Murata_RSI_2008}%
  \BibitemOpen
  \bibfield  {author} {\bibinfo {author} {\bibfnamefont {K.}~\bibnamefont
  {Murata}}, \bibinfo {author} {\bibfnamefont {K.}~\bibnamefont {Yokogawa}},
  \bibinfo {author} {\bibfnamefont {H.}~\bibnamefont {Yoshino}}, \bibinfo
  {author} {\bibfnamefont {S.}~\bibnamefont {Klotz}}, \bibinfo {author}
  {\bibfnamefont {P.}~\bibnamefont {Munsch}}, \bibinfo {author} {\bibfnamefont
  {A.}~\bibnamefont {Irizawa}}, \bibinfo {author} {\bibfnamefont
  {M.}~\bibnamefont {Nishiyama}}, \bibinfo {author} {\bibfnamefont
  {K.}~\bibnamefont {Iizuka}}, \bibinfo {author} {\bibfnamefont
  {T.}~\bibnamefont {Nanba}}, \bibinfo {author} {\bibfnamefont
  {T.}~\bibnamefont {Okada}}, \bibinfo {author} {\bibfnamefont
  {Y.}~\bibnamefont {Shiraga}}, \ and\ \bibinfo {author} {\bibfnamefont
  {S.}~\bibnamefont {Aoyama}},\ }\href@noop {} {\bibfield  {journal} {\bibinfo
  {journal} {Rev. Sci. Instrum.}\ }\textbf {\bibinfo {volume} {79}},\ \bibinfo
  {pages} {085101} (\bibinfo {year} {2008})}\BibitemShut {NoStop}%
\bibitem [{\citenamefont {Eiling}\ and\ \citenamefont
  {Schilling}(1981)}]{A.Eiling_JPFMP_1981}%
  \BibitemOpen
  \bibfield  {author} {\bibinfo {author} {\bibfnamefont {A.}~\bibnamefont
  {Eiling}}\ and\ \bibinfo {author} {\bibfnamefont {J.~S.}\ \bibnamefont
  {Schilling}},\ }\href@noop {} {\bibfield  {journal} {\bibinfo  {journal} {J.
  Phys. F: Met. Phys.}\ }\textbf {\bibinfo {volume} {11}},\ \bibinfo {pages}
  {623} (\bibinfo {year} {1981})}\BibitemShut {NoStop}%
\bibitem [{\citenamefont {Bireckoven}\ and\ \citenamefont
  {Wittig}(1988)}]{B.Bireckoven_JPESI_1988}%
  \BibitemOpen
  \bibfield  {author} {\bibinfo {author} {\bibfnamefont {B.}~\bibnamefont
  {Bireckoven}}\ and\ \bibinfo {author} {\bibfnamefont {J.}~\bibnamefont
  {Wittig}},\ }\href@noop {} {\bibfield  {journal} {\bibinfo  {journal} {J.
  Phys. E: Sci. Instrum.}\ }\textbf {\bibinfo {volume} {21}},\ \bibinfo {pages}
  {841} (\bibinfo {year} {1988})}\BibitemShut {NoStop}%
\bibitem [{sup({\natexlab{a}})}]{sup}%
  \BibitemOpen
  \href@noop {} {} \ \bibinfo {note} {see Supplemental
  Material at xxx for the time dependence of recovery curves.}\BibitemShut
  {Stop}%
\bibitem [{\citenamefont {Kitagawa}\ \emph {et~al.}(2008)\citenamefont
  {Kitagawa}, \citenamefont {Katayama}, \citenamefont {Ohgushi}, \citenamefont
  {Yoshida},\ and\ \citenamefont {Takigawa}}]{K.Kitagawa_JPSJ_2008}%
  \BibitemOpen
  \bibfield  {author} {\bibinfo {author} {\bibfnamefont {K.}~\bibnamefont
  {Kitagawa}}, \bibinfo {author} {\bibfnamefont {N.}~\bibnamefont {Katayama}},
  \bibinfo {author} {\bibfnamefont {K.}~\bibnamefont {Ohgushi}}, \bibinfo
  {author} {\bibfnamefont {M.}~\bibnamefont {Yoshida}}, \ and\ \bibinfo
  {author} {\bibfnamefont {M.}~\bibnamefont {Takigawa}},\ }\href@noop {}
  {\bibfield  {journal} {\bibinfo  {journal} {J. Phys. Soc. Jpn.}\ }\textbf
  {\bibinfo {volume} {77}},\ \bibinfo {pages} {114709} (\bibinfo {year}
  {2008})}\BibitemShut {NoStop}%
\bibitem [{\citenamefont {Kitagawa}\ \emph {et~al.}(2010)\citenamefont
  {Kitagawa}, \citenamefont {Nakai}, \citenamefont {T.Iye}, \citenamefont
  {Ishida}, \citenamefont {Kamihara}, \citenamefont {Hirano},\ and\
  \citenamefont {Hosono}}]{S.Kitagawa_PRB_2010}%
  \BibitemOpen
  \bibfield  {author} {\bibinfo {author} {\bibfnamefont {S.}~\bibnamefont
  {Kitagawa}}, \bibinfo {author} {\bibfnamefont {Y.}~\bibnamefont {Nakai}},
  \bibinfo {author} {\bibnamefont {T.Iye}}, \bibinfo {author} {\bibfnamefont
  {K.}~\bibnamefont {Ishida}}, \bibinfo {author} {\bibfnamefont
  {Y.}~\bibnamefont {Kamihara}}, \bibinfo {author} {\bibfnamefont
  {M.}~\bibnamefont {Hirano}}, \ and\ \bibinfo {author} {\bibfnamefont
  {H.}~\bibnamefont {Hosono}},\ }\href@noop {} {\bibfield  {journal} {\bibinfo
  {journal} {Phys. Rev. B}\ }\textbf {\bibinfo {volume} {81}},\ \bibinfo
  {pages} {212502} (\bibinfo {year} {2010})}\BibitemShut {NoStop}%
\bibitem [{\citenamefont {Nakai}\ \emph {et~al.}(2012)\citenamefont {Nakai},
  \citenamefont {Kitagawa}, \citenamefont {Iye}, \citenamefont {Ishida},
  \citenamefont {Kamihara}, \citenamefont {Hirano},\ and\ \citenamefont
  {Hosono}}]{Y.Nakai_PRB_2012}%
  \BibitemOpen
  \bibfield  {author} {\bibinfo {author} {\bibfnamefont {Y.}~\bibnamefont
  {Nakai}}, \bibinfo {author} {\bibfnamefont {S.}~\bibnamefont {Kitagawa}},
  \bibinfo {author} {\bibfnamefont {T.}~\bibnamefont {Iye}}, \bibinfo {author}
  {\bibfnamefont {K.}~\bibnamefont {Ishida}}, \bibinfo {author} {\bibfnamefont
  {Y.}~\bibnamefont {Kamihara}}, \bibinfo {author} {\bibfnamefont
  {M.}~\bibnamefont {Hirano}}, \ and\ \bibinfo {author} {\bibfnamefont
  {H.}~\bibnamefont {Hosono}},\ }\href@noop {} {\bibfield  {journal} {\bibinfo
  {journal} {Phys. Rev. B}\ }\textbf {\bibinfo {volume} {85}},\ \bibinfo
  {pages} {134408} (\bibinfo {year} {2012})}\BibitemShut {NoStop}%
\bibitem [{\citenamefont {Hirano}\ \emph {et~al.}(2012)\citenamefont {Hirano},
  \citenamefont {Yamada}, \citenamefont {Saito}, \citenamefont {Nagashima},
  \citenamefont {Konishi}, \citenamefont {Toriyama}, \citenamefont {Ohta},
  \citenamefont {Fukazawa}, \citenamefont {Kohori}, \citenamefont {Furukawa},
  \citenamefont {Kihou}, \citenamefont {Lee}, \citenamefont {Iyo},\ and\
  \citenamefont {Eisaki}}]{M.Hirano_JPSJ_2012}%
  \BibitemOpen
  \bibfield  {author} {\bibinfo {author} {\bibfnamefont {M.}~\bibnamefont
  {Hirano}}, \bibinfo {author} {\bibfnamefont {Y.}~\bibnamefont {Yamada}},
  \bibinfo {author} {\bibfnamefont {T.}~\bibnamefont {Saito}}, \bibinfo
  {author} {\bibfnamefont {R.}~\bibnamefont {Nagashima}}, \bibinfo {author}
  {\bibfnamefont {T.}~\bibnamefont {Konishi}}, \bibinfo {author} {\bibfnamefont
  {T.}~\bibnamefont {Toriyama}}, \bibinfo {author} {\bibfnamefont
  {Y.}~\bibnamefont {Ohta}}, \bibinfo {author} {\bibfnamefont {H.}~\bibnamefont
  {Fukazawa}}, \bibinfo {author} {\bibfnamefont {Y.}~\bibnamefont {Kohori}},
  \bibinfo {author} {\bibfnamefont {Y.}~\bibnamefont {Furukawa}}, \bibinfo
  {author} {\bibfnamefont {K.}~\bibnamefont {Kihou}}, \bibinfo {author}
  {\bibfnamefont {C.-H.}\ \bibnamefont {Lee}}, \bibinfo {author} {\bibfnamefont
  {A.}~\bibnamefont {Iyo}}, \ and\ \bibinfo {author} {\bibfnamefont
  {H.}~\bibnamefont {Eisaki}},\ }\href@noop {} {\bibfield  {journal} {\bibinfo
  {journal} {J. Phys. Soc. Jpn.}\ }\textbf {\bibinfo {volume} {81}},\ \bibinfo
  {pages} {054704} (\bibinfo {year} {2012})}\BibitemShut {NoStop}%
\bibitem [{\citenamefont {Zhang}\ \emph {et~al.}(2018)\citenamefont {Zhang},
  \citenamefont {Dmytriieva}, \citenamefont {Molatta}, \citenamefont
  {Wosnitza}, \citenamefont {Khim}, \citenamefont {Gass}, \citenamefont
  {Wolter}, \citenamefont {Wurmehl}, \citenamefont {Grafe},\ and\ \citenamefont
  {K{\"u}hne}}]{Z.T.Zhang_PRB_2018}%
  \BibitemOpen
  \bibfield  {author} {\bibinfo {author} {\bibfnamefont {Z.~T.}\ \bibnamefont
  {Zhang}}, \bibinfo {author} {\bibfnamefont {D.}~\bibnamefont {Dmytriieva}},
  \bibinfo {author} {\bibfnamefont {S.}~\bibnamefont {Molatta}}, \bibinfo
  {author} {\bibfnamefont {J.}~\bibnamefont {Wosnitza}}, \bibinfo {author}
  {\bibfnamefont {S.}~\bibnamefont {Khim}}, \bibinfo {author} {\bibfnamefont
  {S.}~\bibnamefont {Gass}}, \bibinfo {author} {\bibfnamefont {A.~U.~B.}\
  \bibnamefont {Wolter}}, \bibinfo {author} {\bibfnamefont {S.}~\bibnamefont
  {Wurmehl}}, \bibinfo {author} {\bibfnamefont {H.-J.}\ \bibnamefont {Grafe}},
  \ and\ \bibinfo {author} {\bibfnamefont {H.}~\bibnamefont {K{\"u}hne}},\
  }\href@noop {} {\bibfield  {journal} {\bibinfo  {journal} {Phys. Rev. B}\
  }\textbf {\bibinfo {volume} {97}},\ \bibinfo {pages} {115110} (\bibinfo
  {year} {2018})}\BibitemShut {NoStop}%
\bibitem [{\citenamefont {Iye}\ \emph {et~al.}(2012)\citenamefont {Iye},
  \citenamefont {Nakai}, \citenamefont {Kitagawa}, \citenamefont {Ishida},
  \citenamefont {Kasahara}, \citenamefont {Shibauchi}, \citenamefont
  {Matsuda},\ and\ \citenamefont {Terashima}}]{T.Iye_JPSJ_2012}%
  \BibitemOpen
  \bibfield  {author} {\bibinfo {author} {\bibfnamefont {T.}~\bibnamefont
  {Iye}}, \bibinfo {author} {\bibfnamefont {Y.}~\bibnamefont {Nakai}}, \bibinfo
  {author} {\bibfnamefont {S.}~\bibnamefont {Kitagawa}}, \bibinfo {author}
  {\bibfnamefont {K.}~\bibnamefont {Ishida}}, \bibinfo {author} {\bibfnamefont
  {S.}~\bibnamefont {Kasahara}}, \bibinfo {author} {\bibfnamefont
  {T.}~\bibnamefont {Shibauchi}}, \bibinfo {author} {\bibfnamefont
  {Y.}~\bibnamefont {Matsuda}}, \ and\ \bibinfo {author} {\bibfnamefont
  {T.}~\bibnamefont {Terashima}},\ }\href@noop {} {\bibfield  {journal}
  {\bibinfo  {journal} {J. Phys. Soc. Jpn.}\ }\textbf {\bibinfo {volume}
  {81}},\ \bibinfo {pages} {033701} (\bibinfo {year} {2012})}\BibitemShut
  {NoStop}%
\bibitem [{\citenamefont {Dioguardi}\ \emph {et~al.}(2016)\citenamefont
  {Dioguardi}, \citenamefont {Kissikov}, \citenamefont {Lin}, \citenamefont
  {Shirer}, \citenamefont {Lawson}, \citenamefont {Grafe}, \citenamefont {Chu},
  \citenamefont {Fisher}, \citenamefont {Fernandes},\ and\ \citenamefont
  {Curro}}]{A.P.Dioguardi_PRL_2016}%
  \BibitemOpen
  \bibfield  {author} {\bibinfo {author} {\bibfnamefont {A.~P.}\ \bibnamefont
  {Dioguardi}}, \bibinfo {author} {\bibfnamefont {T.}~\bibnamefont {Kissikov}},
  \bibinfo {author} {\bibfnamefont {C.~H.}\ \bibnamefont {Lin}}, \bibinfo
  {author} {\bibfnamefont {K.~R.}\ \bibnamefont {Shirer}}, \bibinfo {author}
  {\bibfnamefont {M.~M.}\ \bibnamefont {Lawson}}, \bibinfo {author}
  {\bibfnamefont {H.-J.}\ \bibnamefont {Grafe}}, \bibinfo {author}
  {\bibfnamefont {J.-H.}\ \bibnamefont {Chu}}, \bibinfo {author} {\bibfnamefont
  {I.~R.}\ \bibnamefont {Fisher}}, \bibinfo {author} {\bibfnamefont {R.~M.}\
  \bibnamefont {Fernandes}}, \ and\ \bibinfo {author} {\bibfnamefont {N.~J.}\
  \bibnamefont {Curro}},\ }\href@noop {} {\bibfield  {journal} {\bibinfo
  {journal} {Phys. Rev. Lett.}\ }\textbf {\bibinfo {volume} {116}},\ \bibinfo
  {pages} {107202} (\bibinfo {year} {2016})}\BibitemShut {NoStop}%
\bibitem [{sup({\natexlab{b}})}]{sup2}%
  \BibitemOpen
  \href@noop {} {} \ \bibinfo {note} {see Supplemental
  Material at xxx for the $\theta$ dependence of $T_{\rm c}$ normalized by
  maximum $T_{\rm c}$ of each system in various iron-based superconductors,
  which includes Refs. [5,8-11,30].}\BibitemShut {Stop}%
\end{thebibliography}
%

\end{document}